\documentclass{PoS}
\usepackage{amsmath}

\title{Extending the eigCG algorithm to non-symmetric linear systems with multiple
right-hand sides}
\ShortTitle{Extending the eigCG algorithm to non-symmetric linear systems with
multiple right-hand sides}
\author{\speaker{Abdou Abdel-Rehim}\\%
        Department of Physics and Department of Computer Science,\\
        The College of William \& Mary, Williamsburg, VA, U.S.A. and\\
        Thomas Jefferson National Accelerator Facility, Newport News, VA, U.S.A.\\
        E-mail: \email{amrehim@cs.wm.edu}}
\author{Kostas Orginos\\
        Department of Physics, The College of
        William \& Mary, Williamsburg, VA, U.S.A. and\\
        Thomas Jefferson National Accelerator Facility, Newport News, VA, U.S.A.\\
        E-mail: \email{kostas@jlab.org}}

\author{Andreas Stathopoulos\\
        Department of Computer Science, The College of William \& Mary, Williamsburg, VA, U.S.A.\\
        E-mail: \email{andreas@cs.wm.edu}} 

\abstract{ For Hermitian positive definite linear systems and eigenvalue problems, the eigCG
          algorithm is a memory efficient algorithm that solves the linear system and
          simultaneously computes some of its eigenvalues. The algorithm is
          based on the Conjugate-Gradient (CG) algorithm, however, it uses only a
          window of the vectors generated by the CG algorithm to compute approximate
          eigenvalues. The number and accuracy of the eigenvectors can be increased
          by solving more right-hand sides. For Hermitian systems with multiple
          right-hand sides, the computed eigenvectors can be used to speed
          up the solution of subsequent systems. The algorithm was tested on Lattice QCD
          problems by solving the normal equations and was shown to give large speed up
          factors and to remove the critical slowing down as we approach light quark masses.
          Here, an extension to the non-symmetric case based on the two-sided Lanczos algorithm is
          given. The new algorithm is tested on Lattice QCD problems and is shown to give very
          promising results. We also study the removal of the critical slowing down and 
          compare results with those of the eigCG algorithm. We also discuss the case when
          the system is $\gamma_5$-Hermitian.}

\FullConference{The XXVII International Symposium on Lattice Field Theory - LAT2009\\
		 July 26-31 2009\\
		 Peking University, Beijing, China}

\begin{document}

\section{Introduction}
\label{sect:introduction}
Computation of various hadronic properties from Lattice QCD requires evaluation
of the contribution of disconnected quark loops. This include, for example, the 
mass of the neutral pion, the spectrum of Isospin singlet mesons\cite{Gregory:2007ev}, and the contribution
of strange sea quarks to the electromagnetic form factors of the proton\cite{Leinweber:2006ug},\cite{Lewis:2002ix}.  
Evaluation of the contribution of disconnected quark loops requires knowledge of the quark propagator from all
sites to all sites on the lattice (all-to-all propagators)\cite{Wilcox:1999ab,Foley:2005ac,Peardon:2009gh}. 
In all-to-all propagator methods, one is required to compute the action of the
inverse of the lattice Dirac operator $A$ on a particular set of sources $b_i, \quad i=1,2,\dots,N_r$, by solving
the linear systems,
\begin{equation}
A x_i = b_i, \quad i=1,2,\dots,N_r. 
\label{main-problem}
\end{equation}
Typical values of $N_r$ are ${\cal{O}}(50-100)$ and large values of $N_r$ are required for smaller statistical noise errors.
In addition, for small quark masses, solving Eqs.\ref{main-problem} using standard 
iterative methods, such as GMRES or BiCGStab, converges very slowly (critical 
slowing down phenomena). It has been realized that critical slowing down could be removed by computing and deflating the lowest
eigenmodes of $A$ (see \cite{Wilcox:2007ei} for a review of deflation methods in lattice QCD).

In \cite{Stathopoulos:2007zi} we have given the Incremental eigCG algorithm for Hermitian,
positive definite systems. The eigCG part of the algorithm
solves a single system using the Conjugate Gradient (CG) algorithm and
simultaneously computes few eigenvectors with smallest eigenvalues. EigCG uses only a small size window 
of the CG residuals for computing eigenvalues. In addition,
the standard CG part of eigCG for solving the linear system is totally unaffected by the computation of the 
eigenvalues. For multiple right-hand sides, Incremental eigCG solves a small subset of the linear systems using
eigCG and concurrently accumulates more eigenvectors as desired. The remaining systems are then solved with CG
after deflating the computed eigenvectors from the initial guesses. Incremental eigCG was tested on large Lattice QCD 
problems \cite{Stathopoulos:2007zi} with very small quark masses and was found to remove the critical slowing 
down as well as speed up the solution for multiple right-hand sides through deflation. Since the Dirac matrix $A$ is 
non-Hermitian, it was necessary to apply Incremental eigCG to the normal equations,
\begin{equation}
A^\dagger A x_i = A^\dagger b_i, \quad i=1,2,\dots,N_r. 
\label{normal-equations-problem}
\end{equation}
In this report, we present and test an extension of the ideas of eigCG and its incremental version to the 
non-Hermitian case. There are three motivations for studying this extension. First, converting the non-Hermitian
system \ref{main-problem} into the Hermitian, positive definite system \ref{normal-equations-problem} leads to a more
difficult system as the new system will have a worse condition number. Second, solving the non-Hermitian system will give
eigenvalues of $A$ directly which could be useful for other applications. Finally, one would like to compare
the efficiency of removing the critical slowing down when solving the systems \ref{main-problem} and \ref{normal-equations-problem}.
In the extension to non-Hermitian case, we first add functionality to the BiCG algorithm, following closely what was done 
in the Hermitian case, that allows for computing few eigenvalues using only a limited size window of the BiCG residuals. In this case
we'll need to compute left and right eigenvectors of $A$. The modified BiCG algorithm will be called eigBiCG. 
For multiple right-hand sides, we solve a subset of the systems using eigBiCG and accumulate more eigenvectors 
and, hopefully, improve their accuracy using an incremental scheme as was done in the Hermitian case. For the 
remaining systems, we deflate the components of the
computed eigenvectors and then use BiCGStab to solve them. Using BiCGStab instead of BiCG is motivated by the fact that
BiCGStab normally converges faster than BiCG. We chose BiCG for computing eigenvectors because the BiCG residuals
and parameters can easily be related to the Bi-Lanczos vectors and projection matrix. We also discuss simplifications 
when $A$ satisfies the $\gamma_5$-Hermiticity condition $\gamma_5 A=A^\dagger \gamma_5$.

In the following, the dot product of two vectors will be denoted by $(w,v):=w^\dagger v$, the Euclidean norm of a vector is
denoted by $||v||$ and the complex conjugate of a number $z$ will be denoted by $\bar{z}$. The function $[zr,zl,D]=eig(C)$ returns the right 
eigenvectors $zr$, the left eigenvectors $zl$ and the eigenvalues array $D$ sorted according to a user chosen criteria.
\section{Incremental eigBiCG algorithm}
\label{sect:incr-eigbicg}
In the BiLanczos algorithm one solves the dual systems $Ax=b$ and
$A^\dagger \hat{x}=\hat{b}$. Given $x_0$ and $\hat{x}_0$ initial guesses, the algorithm builds a 
bi-orthogonal basis for the Krylov subspaces,
\begin{eqnarray}
{\cal{K}}(A,v_1)&=&\{v_1,Av_1,A^2v_1,\dots\} \\ \nonumber
\hat{{\cal{K}}}(A^\dagger,w_1) & = & \{w_1,A^\dagger w_1,A^{\dagger 2} w_1, \dots\}, 
\label{eq:bi-lanz-krylov}
\end{eqnarray}    
where $v_1=r_0/||r_0||, \ r_0=b-Ax_0,$ 
\ $\hat{r}_0=\hat{b}-A^\dagger \hat{x}_0$ 
and $w_1 = \hat{r}_0 /(r_0^\dagger \hat{r}_0)$, so that 
$w^\dagger_1 v_1=1$. 
Normally, $\hat{b}=b,\ \hat{x}_0=x_0,$ and $w_1=v_1$. Let $V^{(m)}=\{v_1,v_2,\dots,v_m\}, \quad W^{(m)}=\{w_1,w_2,\dots,w_m\}$ be 
the bi-orthogonal bases of ${\cal{K}},$ and $\hat{{\cal{K}}}$, and $H^{(m)}=W^{(m) \dagger} A V^{(m)}$ the
projection matrix. Let $y^{(m)}$ and $z^{(m)}$ be the right and left eigenvectors of $H^{(m)}$. The approximate
right and left Ritz eigenvectors of $A$ are given by $Y^{(m)}=V^{(m)}y^{(m)}$ and $Z^{(m)}=W^{(m)}z^{(m)}$ 
respectively. 
In the BiCG algorithm, the bi-orthogonal bases $V,\ W$ and the projection 
matrix $H$ are not computed explicitly. 
The basis vectors are obtained as the right and left BiCG residuals, while
the matrix $H$ is obtained from the BiCG scalar coefficients. This can be done by noting that
$v_j=\eta_j r_{j-1}$ and $w_j=\zeta_j \hat{r}_{j-1}$ for $j=1,2,\dots$, with 
$\eta_j$ and $\zeta_j$ satisfying
$(w_j,v_j)=1$. In the following, we choose a normalization such that $||v_j||=1$ and $\eta_j$ is real 
positive, however, other normalizations that maintain
the biorthogonality of $w_j$ and $v_j$ could be used. From these relations, and the biorthogonality of the BiCG
residuals, the elements of $H$ could be recovered from the scalar coefficients of BiCG without extra matrix-vector products.

Similar to eigCG, the eigBiCG algorithm adds functionality to the standard BiCG algorithm for computing few
eigenvalues and eigenvectors of $A$ using only a window of size $m$ of the BiCG residuals. For $nev$ requested
eigenvalues with a chosen criterion (smallest absolute value, for example), the eigenvalue part
of eigBiCG computes eigenvectors from the size $m$ and size $(m-1)$ subspaces.
The search subspaces $V^{(m)},\quad W^{(m)}$ and the projection matrix $H^{(m)}$ are then restarted
with these $2nev$ computed eigenvectors which are inexpensively biorthogonalized in the coefficient space. The details of this part 
are given in Algorithm $BiCG-eigen$. Note that we need $m > 2nev$.
The full eigBiCG algorithm for solving the linear system $Ax=b$ and computing $nev$ eigenvalues and
eigenvectors using a search subspace of dimension $m$ is given in Algorithm $eigBiCG$. In order
to compute the elements of the $2nev+1$-th row and column of $H$ after restarting we need $Av_{2nev+1}$
and $A^\dagger w_{2nev+1}$ which will be given from $Ar_{j-1}$ and $A^\dagger \hat{r}_{j-1}$. This can be accomplished
by using the relations $p_{j-1}=r_{j-1}+\beta_{j-2}p_{j-2}$ and  $\hat{p}_{j-1} = \hat{r}_{j-1} + \bar{\beta}_{j-2}\hat{p}_{j-2}$.
So, we need to store $Ap$ and $A^\dagger \hat{p}$ products when $vs=m-1$, where $vs$ is the current size of the search subspaces. 
\begin{center}
\begin{minipage}{0.85\textwidth}
{\small
\rule{\textwidth}{0.7mm}
{\bf Algorithm: $BiCG-eigen(nev,m,V^{(m)},W^{(m)},H^{(m)})$}\\
\rule{\textwidth}{0.3mm}
\begin{itemize}
\item $[y^{(m)},z^{(m)},E^{(m)}]=eig(H^{(m)})$, 
$\quad \quad [y^{(m-1)},z^{(m-1)},E^{(m-1)}]=eig(H^{(m-1)})$. 
\item Append a zero at the end of each of the vectors $y^{(m-1)},z^{(m-1)}$.
\item $[\tilde{y},\tilde{z}]=$ Bi-orthogonalize 
                         $[y^{(m)}_1,y^{(m)}_2,\dots,y^{(m)}_{nev};\quad y^{(m-1)}_1,y^{(m-1)}_2,\dots,y^{(m-1)}_{nev}]$ 
against $[z^{(m)}_1,z^{(m)}_2,\dots,z^{(m)}_{nev};\quad z^{(m-1)}_1,z^{(m-1)}_2,\dots,z^{(m-1)}_{nev}]$.
Note that $y^{(m)},z^{(m)}$ are already biorthogonal. Need only to extend biorthogonality to the rest of vectors. 
\item $T=\tilde{z}^\dagger H^{(m)} \tilde{y}$, $\quad \quad [u,q,\Lambda]=eig(T)$.
\item $U=V^{(m)}\tilde{y}u$,\quad $Q=W^{(m)}\tilde{z}q$.
\item {\bf  Restart:}
\begin{itemize}
\item $V^{(m)}=[\ \ ],\quad W^{(m)}=[\ \ ]$, \quad \quad $V^{(m)}_{1:2nev}=U,\quad W^{(m)}_{1:2nev}=Q$.
\item $H^{(m)}_{i,j}=0; \quad i,j=1,2,\dots,m$, \quad \quad  $H^{(m)}_{i,i}=\Lambda_i$ for $i=1,2,..,2nev$.
\end{itemize}
\end{itemize}
\rule{\textwidth}{0.7mm}
}
\end{minipage}
\end{center}
\begin{center}
\begin{minipage}{0.85\textwidth}
{\small
\rule{\textwidth}{0.7mm}
{\bf Algorithm $Incremental \ eigBICG$}\\
\rule{\textwidth}{0.3mm}
Given initial guesses $x^k_0$ for $k=1,2,\dots,N_r$:
\begin{enumerate}
\item Choose $nev,m$ and set $U_l=[\quad],\quad U_r=[\quad]$,
      and $\quad H=[\quad]$.
\item {\bf For $k=1,2,\dots,n_1$}
\begin{itemize}
\item {\bf If $U_r$ is not empty}, set $x^k_0 = x^k_0+U_rd$, where $Hd=U_l^\dagger(b^k-Ax^k_0)$.
\item solve the system using $eigBiCG(nev,m,V,W)$.
\item Compute $[V',W']=$ biorthogonalize $[V,W]$ against $[U_r,U_l]$.
\item Compute the new 
$
             H = \begin{pmatrix}
                 H   &   U_l^\dagger A V' \\
                 W'^\dagger AU_r & W'^\dagger A V' \\
                \end{pmatrix} \nonumber
$
\item Add the new vectors: $U_l=[U_l \quad W']$ and $U_r=[U_r \quad V']$.
\end{itemize}
\item {FOR $k=n_1+1,n_1+2,\dots,N_r$}
\begin{itemize}
\item $x^k_0 = x^k_0+U_rd$, where $Hd=U_l^\dagger (b^k-Ax^k_0)$.
\item Solve the system using BiCGStab.
\item Repeat the deflation and restart BiCGStab when the residual is less than $DefTol*||b||$.
\end{itemize}
\end{enumerate}
\rule{\textwidth}{0.3mm}
}
\end{minipage}
\end{center}

\begin{center}
\begin{minipage}{0.85\textwidth}
{\small
\rule{\textwidth}{0.7mm}
{\bf Algorithm: $eigBiCG(nev,m,\Lambda,U,Q)$}\\
\rule{\textwidth}{0.3mm}
\begin{enumerate}
\item Choose initial guess $x_0$, compute $r_0=b-Ax_0$, and set $p_0=r_0$.
\item Choose $\hat{r}_0$ such that $(\hat{r}_0,r_0) \ne 0$, and set $\hat{p}_0=\hat{r}_0$.
Set $\rho_0=(\hat{r}_0,r_0)$, $vs=0$.
\item {\bf For  $j=1,2,\dots$ till convergence}
\begin{itemize}
\item Compute $Ap_{j-1}$ and $A^\dagger \hat{p}_{j-1}$.
\item Compute $\sigma_{j-1}=(\hat{p}_{j-1},Ap_{j-1})$ and $\alpha_{j-1} = \frac{\rho_{j-1}}{\sigma_{j-1}}$.
Set $x_j = x_{j-1} + \alpha_{j-1} p_{k-1}$.
\item {\bf If $vs=m-1$}, $q= Ap_{j-1}$, $s= A^\dagger \hat{p}_{j-1}$.
\item {\bf If $vs=m$}
\begin{itemize}
\item Compute eigenvalues and restart the search subspace and projection matrix 
       using the algorithm $BiCG-eigen(nev,m,V,W,H)$.
Set $vs=2nev$.
\item Compute the $H_{k,2nev+1}$ and $H_{2nev+1,k}$ for $k=1,2,\dots,2nev$.\\
$H_{2nev+1,k} = \frac{\|r_{j-1}\|}{\rho_{j-1}}(A^\dagger \hat{p}_{j-1}-\bar{\beta}_{j-2} s)^\dagger v_k. \quad \\
H_{k,2nev+1} = \frac{1}{\|r_{j-1}\|}w_k^\dagger(Ap_{j-1}-\beta_{j-2} q).$
\end{itemize}
\item     $vs = vs +1,\quad v_{vs}=\frac{1}{\|r_{j-1}\|}r_{j-1}$,
           $w_{vs}=\frac{\|r_{j-1}\|}{\bar{\rho}_{j-1}}\hat{r}_{j-1}$.
\item Compute $r_j = r_{j-1} - \alpha_{j-1}Ap_{j-1}$ and
       $\hat{r}_j = \hat{r}_{k-1} - \bar{\alpha}_{j-1}A^\dagger \hat{p}_{j-1}$.
\item Set $\rho_j=(\hat{r}_j,r_j)$ and compute $\beta_{j-1}=\frac{\rho_j}{\rho_{j-1}}$.
\item Set $p_j = r_j + \beta_{j-1}p_{j-1}$ and
           $\hat{p}_j = \hat{r}_j + \bar{\beta}_{j-1}\hat{p}_{j-1}$.

\item Compute the diagonal $H$ matrix elements:\\
if $j=1$, $H_{vs,vs}= \frac{1}{\alpha_{j-1}},$ else
          $H_{vs,vs}=\frac{1}{\alpha_{j-1}}+\frac{\beta{j-2}}{\alpha_{j-2}}$.
\item {\bf If $vs<m$}, compute the off-diagonal $H$ matrix elements:\\
$H_{vs,vs+1}= - \frac{\|r_{j-1}\|}{\|r_j\|} \frac{\beta_{j-1}}{\alpha_{j-1}}, \quad 
              H_{vs+1,vs}  =   - \frac{\|r_j\|}{\|r_{j-1}\|}\frac{1}{\alpha_{j-1}}$.
            
\item If $||r_j|| \le tol*||b||$ for a given tolerance $tol$, stop the iterations.
\end{itemize}
\item Using $V^{(vs)}$, $W^{(vs)}$ and $H^{(vs)}$ compute the final $nev$ eigenvalues and eigenvectors:\\
$[y^{(vs)},z^{(vs)},\Lambda^{(vs)}]=eig(H^{(vs)}), \quad \quad U^{(nev)}=V^{(vs)}y^{(nev)}, \quad \quad
Q^{(nev)}=W^{(vs)}z^{(vs)}$.
\end{enumerate}
\rule{\textwidth}{0.3mm}
}
\end{minipage}
\end{center}

For multiple right-hand sides, we use the Incremental eigBiCG algorithm.
After solving a subset of the right-hand sides and accumulating the deflation subspaces $U_l$ and $U_r$, we use 
BiCGStab on the remaining systems after deflating the eigenvector components. Since computed 
eigenvectors are not exact we might need to repeat the deflation step and restart BiCGStab depending on the accuracy
of the eigenvectors. The deflation restart tolerance is called $DefTol$. In addition, final eigenvectors 
computed from incremental eigBiCG could be computed, if necessary,
using Raleigh-Ritz with $U_l$, and $U_r$ as search subspaces.\\

\section{Results}
\begin{figure}[htb]
\begin{center}
\includegraphics[width=0.49\textwidth]{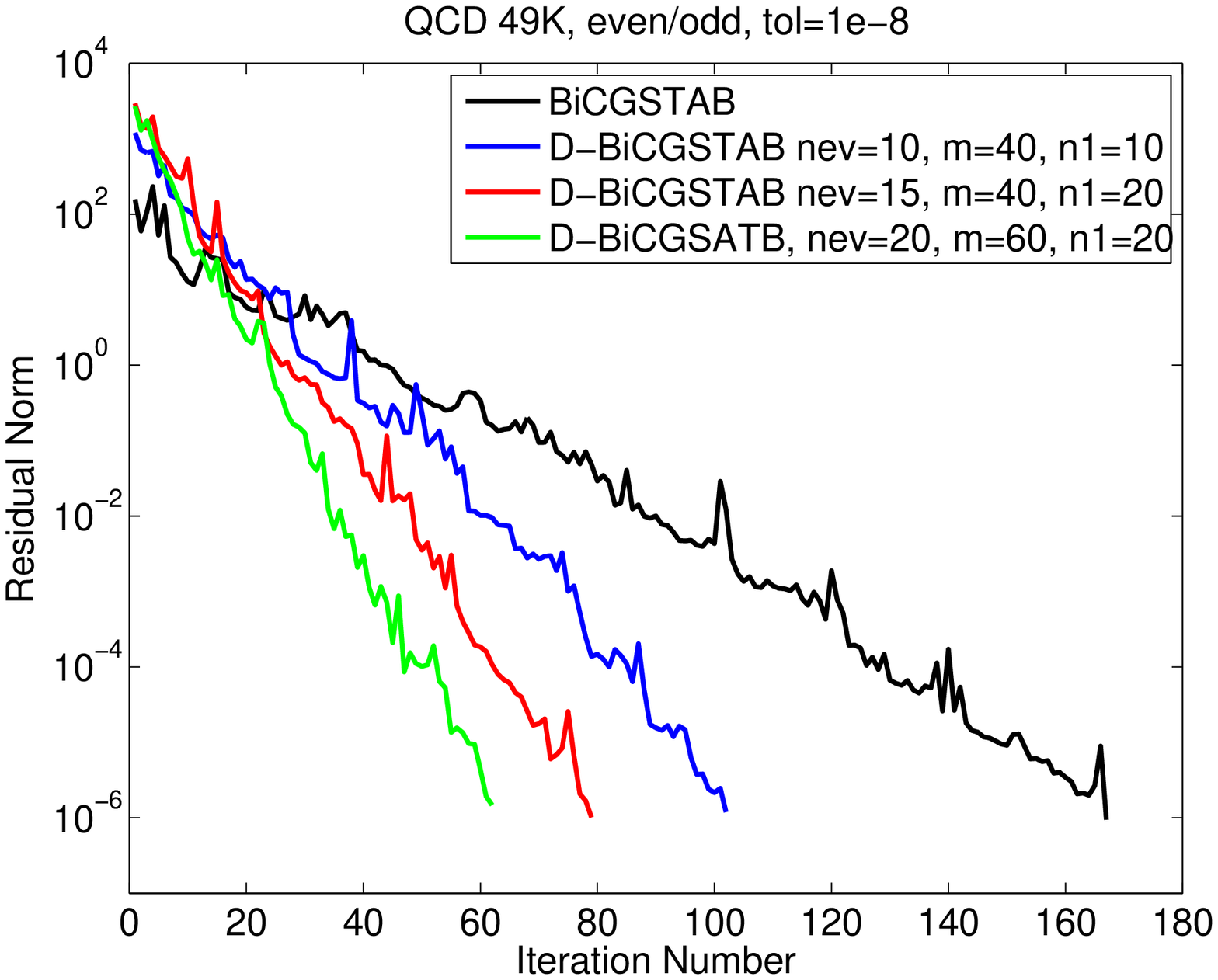}
\includegraphics[width=0.49\textwidth]{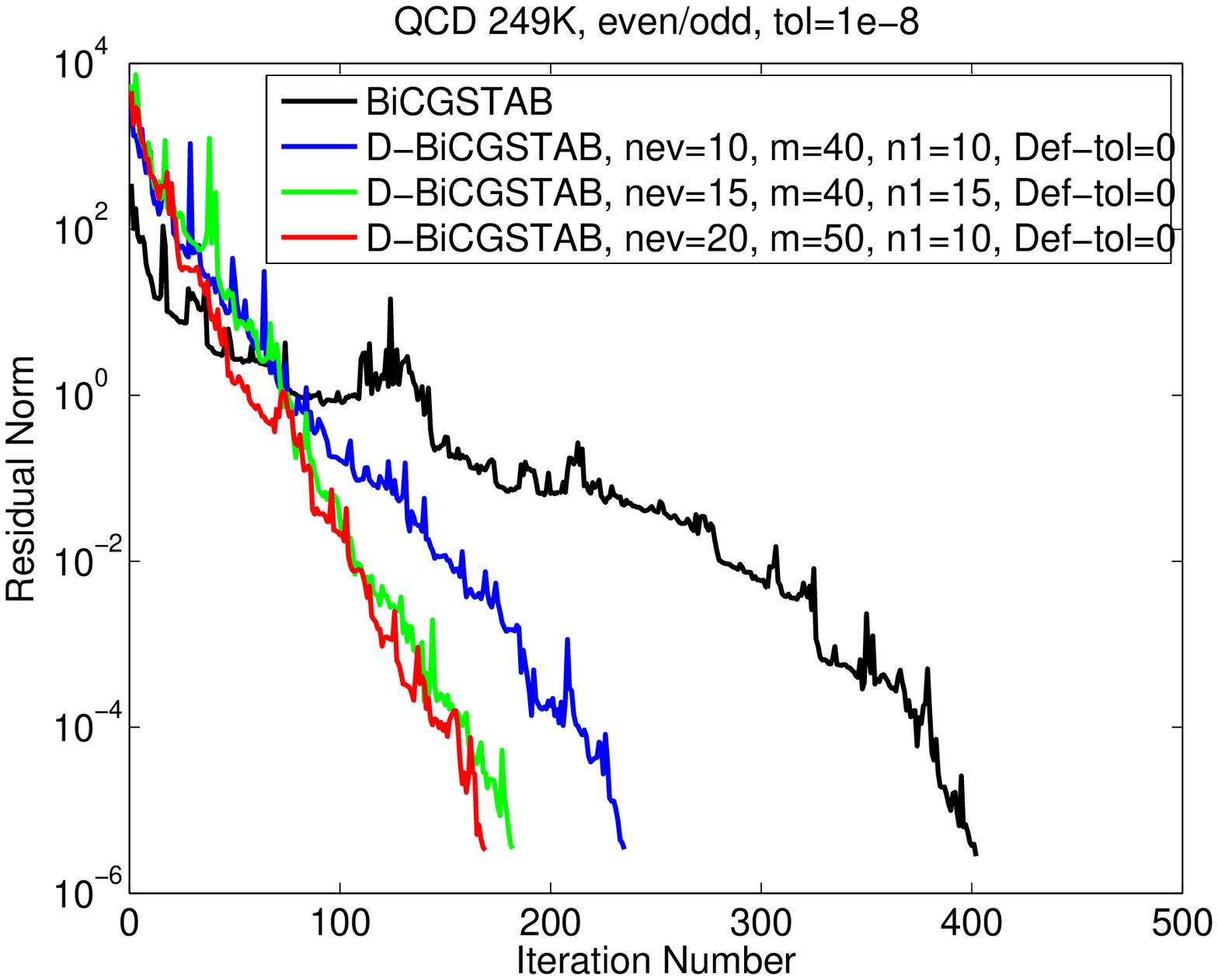}
\caption{Deflated BiCGStab using eigenvectors computed with eigBiCG for $n1$ right-hand sides
for different choices of $nev$ and $m$. No restarting was needed ($DefTol=0$ was used).}
\label{fig:deflation_curves}
\end{center}
\end{figure}

\begin{figure}[htb]
\begin{center}
\includegraphics[width=0.49\textwidth]{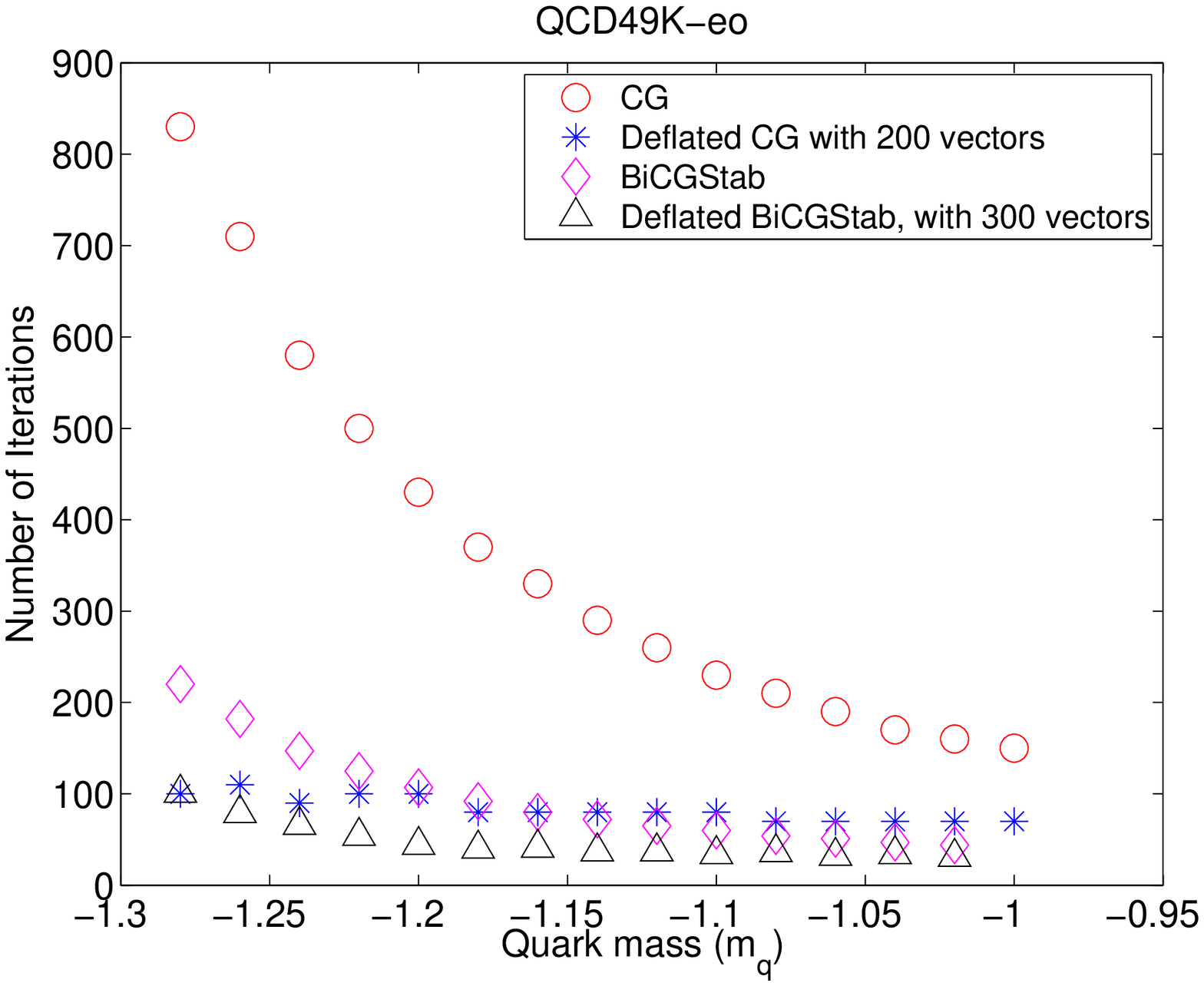} 
\includegraphics[width=0.49\textwidth]{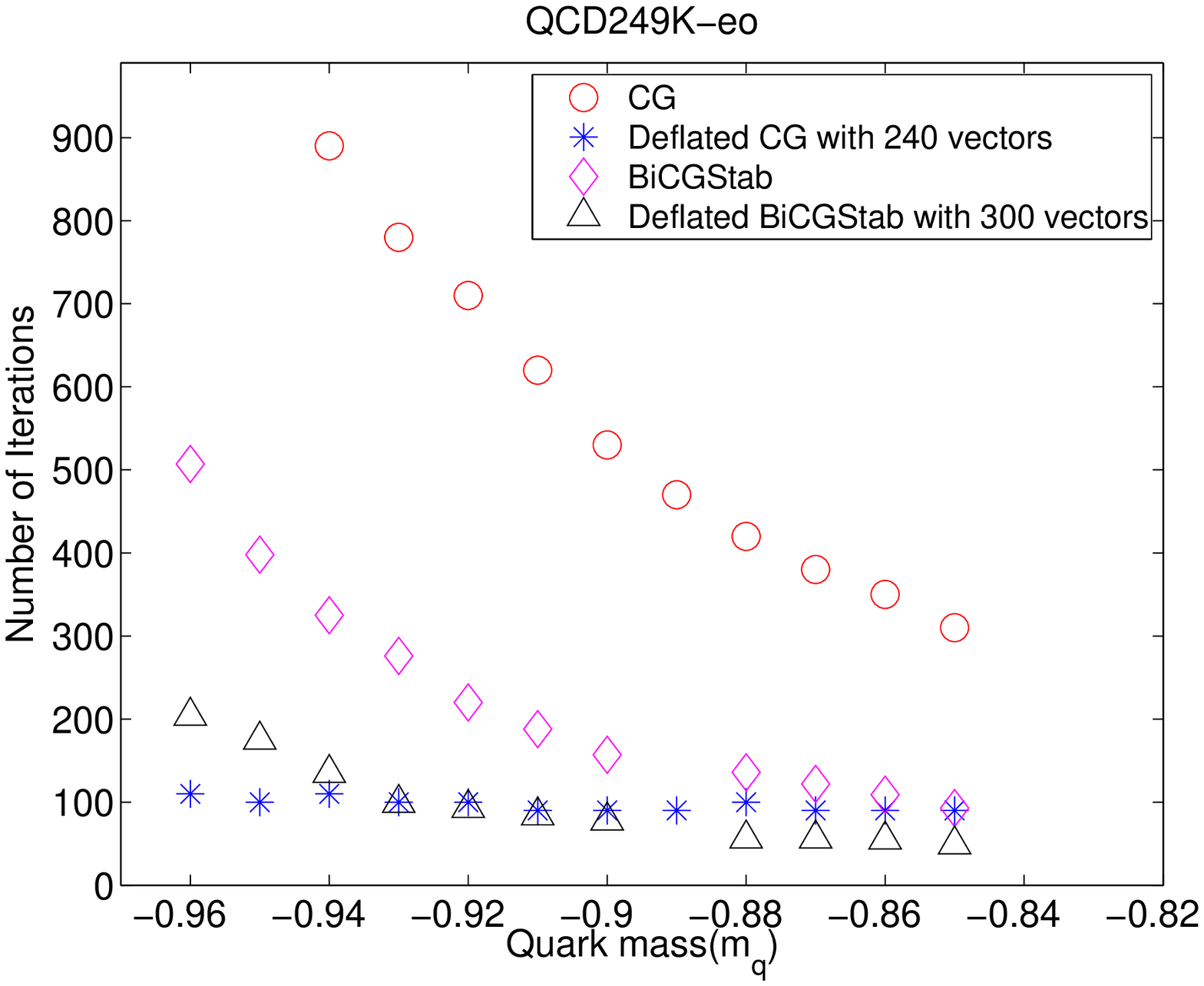}
\caption{Removing critical slowing down using eigBiCG and eigCG applied to the normal equations.}
\label{fig:no_crit_sd}
\end{center}
\end{figure}

The algorithm is preliminary tested on two quenched Wilson lattice QCD matrices with even-odd preconditioning 
near$\kappa_{critical}$. The first is a $8^4$ lattice at $\beta=5.5$ with $m_q=-1.25$ where $\kappa=\frac{1}{8+2m_q}$. 
This case will be labeled as $QCD49K-eo$ since the Dirac matrix will be of size $49,152$ before the even-odd precondtioning.
The second is a $12^4$ lattice at $\beta=5.8$ with $m_q=-0.95$. This case will be labeled as $QCD249K-eo$.
We first compare the lowest eigenvalues computed with eigBiCG to those computed with un-restarted BiCG in which
all the residuals were stored. As seen from Table \ref{Tab:comp-bicg-249k}, the results from eigBiCG with 
a limited storage gives eigenvalues in close agreement with un-restarted BiCG where all the residuals were stored. 
We next study how incremental eigBiCG could speed up the solution with many right-hand sides. In Figure \ref{fig:deflation_curves}, 
we show the effect of deflation for different choices of $nev$ and $m$ after solving $n1$ right-hand sides, showing a speed up factor 
of about $2.5$. In Figure \ref{fig:no_crit_sd}, we show the effect of reducing the quark mass on the number of iterations used by 
$BiCGStab$ and compare it to the case when solving the normal equations using $Incremental \ EigCG$. The results show that 
$Incremental \ eigBiCG$ is competitive with $eigCG$ but not necessarily better. We note that both $BiCGStab$ and $CG$ applied
to the normal equations use two matrix-vector products per iteration. A better comparison between the two methods requires 
experiments on larger lattice QCD matrices.

\begin{table}[htb]
{\small
\begin{center}
\begin{tabular}{|c|c| c| }
\hline
Method &  Eigenvalues    &  Residuals    \\
\hline
eigBiCG               &3.46577e-03-1.07644e-13i& 1.71e-07 \\
                      &1.35450e-02+1.72604e-02i& 1.69e-06 \\
                      &1.35450e-02-1.72604e-02i& 1.37e-06 \\
                      &2.82870e-02+1.09765e-07i& 2.40e-03  \\
                      &1.51950e-02-2.26792e-02i& 9.35e-01 \\
\hline
BiCG         &3.46577e-03+7.84451e-15i& 1.43e-08 \\
             &1.35450e-02+1.72604e-02i& 1.72e-06 \\
             &1.35450e-02-1.72604e-02i& 1.40e-06 \\
             &2.82870e-02+1.09755e-07i& 2.39e-03 \\
             &1.36523e-02+4.16515e-02i& 1.52e-06 \\
\hline
\end{tabular}
\caption{Comparing lowest 5 eigenvalues for QCD249K-eo obtained with un-restarted BiCG and with eigBiCG using $nev=15$ and $m=40$.
The tolerance for the linear system was chosen to be $1e-08$ and the system converged in $592$ iterations}
\label{Tab:comp-bicg-249k}
\end{center}
}
\end{table}

\section{$\gamma_5$-Hermitian systems }
\begin{figure}[htb]
\begin{center}
\includegraphics[width=0.65\textwidth]{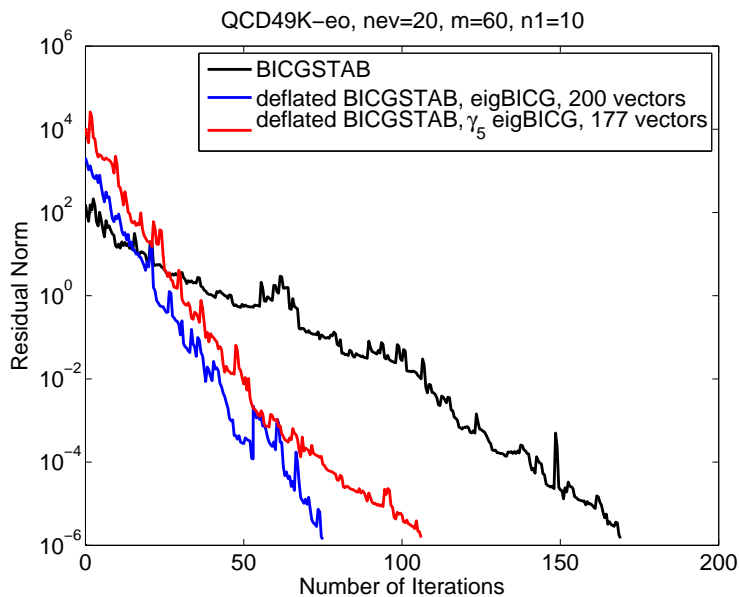} 
\caption{Comparison of deflation with $\gamma_5$-Hermitian algorithm.}
\label{fig:g5_deflation_curves}
\end{center}
\end{figure}

For Wilson and Clover fermions we have the symmetry 
\begin{equation}
\gamma_5 A = A^\dagger \gamma_5. 
\end{equation}
Using this symmetry we can replace the costly matrix-vector multiplication with $A^\dagger$
in eigBiCG with the cheaper multiplication with $\gamma_5$.
In BiCG, if we chose $\hat{r}_0=\gamma_5 r_0$ then it follows that 
$\hat{r}_j=\gamma_5 r_j$ and the search directions $\hat{p}_j=\gamma_5 p_j$ for subsequent iterations.  
Also, eigenvalues will be real or come in pairs of conjugate values, and left eigenvectors 
are computable from right ones, as long as we keep eigenvectors corresponding to conjugate pairs. 
Using these relations, we can simplify the first phase of $Incremental \ eigBiCG$ where eigenvalues
are computed. For illustration, we show preliminary results comparing the two versions of the algorithm
in Figure \ref{fig:g5_deflation_curves}. The result shows a similar performance In which only the right
eigenvectors need to be stored and where matrix-vector multiplication with $A^\dagger$ is avoided.  

\section{Conclusions}
Extending the ideas behind the successful eigCG algorithm to non-Hermitian systems gave very promising
results. The new algorithm gave access to the left and right eigenvectors of the Dirac matrix while
solving the linear systems using only a limited storage. It was also shown to remove the critical slowing down
and to be competitive with eigCG. For $\gamma_5$-Hermitian systems, preliminary study shows that storage
of the left eigenvectors and multiplication with $A^\dagger$ could be avoided. 

\acknowledgments
This work was supported by the National Science Foundation grant
CCF-0728915, the Jefferson Science Associates under U.S. DOE Contract No. DE-AC05-06OR23177 
and the Jeffress Memorial Trust grant J-813.

\end{document}